\documentclass[amssymb,amsmath,pra,twocolumn,floatfix,showpacs,superscriptaddress]{revtex4}
\usepackage{graphicx}

\begin{document}
\title{Quantum gate for Q switching in
monolithic photonic bandgap cavities containing two-level atoms.}

\author{Andrew D.~Greentree}
\email{andrew.greentree@ph.unimelb.edu.au} \affiliation{Centre for
Quantum Computer Technology, School of Physics, The University of
Melbourne, Melbourne, Victoria 3010, Australia.}

\author{J. Salzman}
\affiliation{Microelectronics Research Center, Electrical
Engineering, Technion, Haifa 32000, Israel.}

\author{Steven Prawer}
\affiliation{Centre for Quantum Computer Technology, School of
Physics, The University of Melbourne, Melbourne, Victoria 3010,
Australia.}

\author{Lloyd C. L. Hollenberg}
\affiliation{Centre for Quantum Computer Technology, School of
Physics, The University of Melbourne, Melbourne, Victoria 3010,
Australia.}

\date{November 10, 2005}

\begin{abstract}
Photonic bandgap cavities are prime solid-state systems to
investigate light-matter interactions in the strong coupling regime.
However, as the cavity is defined by the geometry of the periodic
dielectric pattern, cavity control in a monolithic structure can be
problematic. Thus, either the state coherence is limited by the
read-out channel, or in a high Q cavity, it is nearly decoupled from
the external world, making measurement of the state extremely
challenging.  We present here a method for ameliorating these
difficulties by using a coupled cavity arrangement, where one cavity
acts as a switch for the other cavity, tuned by control of the
atomic transition.
\end{abstract}

\pacs{42.50.Pq, 42.70.Qs, 42.60.Gd, 32.80.-t}

\maketitle

%%%%%%%%%%%%%%%%%%%%%%%%%%%%%%%%%%%%%%%%%%%%%%%%%%%%%%%%%%%%%%%%

As the maturity and sophistication of quantum optics progresses,
there is a growing movement to translate such effects into practical
devices.  This impetus suggests, for reasons of scalability and
practicality, the need for viable solid-state technologies to
produce and distribute single photons as an enabling technology for
derivative quantum devices.  In particular we are concerned with the
role played by cavity Quantum Electro-Dynamics (CQED) in such
devices.

CQED has been used to great effect in the generation of
deterministic, transform limited single (and higher-order Fock
states) photon pulses \cite{bib:SinglePhoton}, and schemes exist
which incorporate CQED for quantum computing \cite{bib:CQEDQC}, and
entanglement generation \cite{bib:Entangled}. More recently `hybrid'
schemes for quantum computation have been suggested incorporating
matter qubits in cavities with single photon generation, linear
optics and high fidelity photon detection \cite{bib:Hybrid}. However
many of these schemes (with notable exceptions) will be problematic
to scale or to remove from laboratory environments.

Given difficulties with implementing most present schemes in
non-research environments, significant attention has turned towards
photonic band gap (PBG) cavities as quantum cavities.  This is due
to their superb photonic confinement properties and the recent
realization of high Q cavities with small mode volume (of order the
wavelength$^3$) \cite{bib:PBCav,bib:SongNM2005}.  These successes
have been fueled by a combination of technological imperatives and
advances in fabrication.

A PBG material is created by producing a periodic modulation in the
dielectric function of a material so that Bragg interference
prevents propagation of certain modes across the structure. Such
structures may be two-dimensional, with confinement in the third
dimension realized by classical waveguiding, or by creating a
three-dimensional lattice.  We concentrate on the former example as
it is easier to produce, and has so far yielded the most dramatic
effects. The most popular configuration for 2D PBG structures is a
thin membrane with a 2D array of holes (a lattice) drilled in it.  A
defect (usually an undrilled hole, local variation in lattice
spacing, or combination of both) defines a PBG cavity, as any photon
injected into that site cannot propagate laterally away from the
defect. In this way, PBG cavities can constitute extremely good
cavities with low loss, high coupling and low mode volume, all
necessary conditions for probing the strong-coupling limit of CQED.

One problem with high-Q cavities is the difficulty of out-coupling
excitations from the cavity \cite{bib:Khanbekyan2004}.  One would
like a Q-switch, a device that can be modulated in some fashion to
change the cavity from high Q to low Q, with the optical intensity
dumped from the cavity in a controlled fashion: we will term such a
device a ``gate".  Q-switching is well-known for classical laser
applications \cite{bib:YarivBook}, but is less easy for PBG
cavities, although some recent proposals exist including mechanical
switches \cite{bib:Koenderink2004}, and nonlinear optical effects
\cite{bib:NonlinearOptical}. However in monolithic structures where
we cannot use mechanical or thermal effects, and operating at low
light levels, there have been no suitable suggestions for an
effective Q-switch in PBG cavities. This is the problem addressed in
this paper.

\begin{figure}[tb!]
\includegraphics[width=1.0\columnwidth,clip]{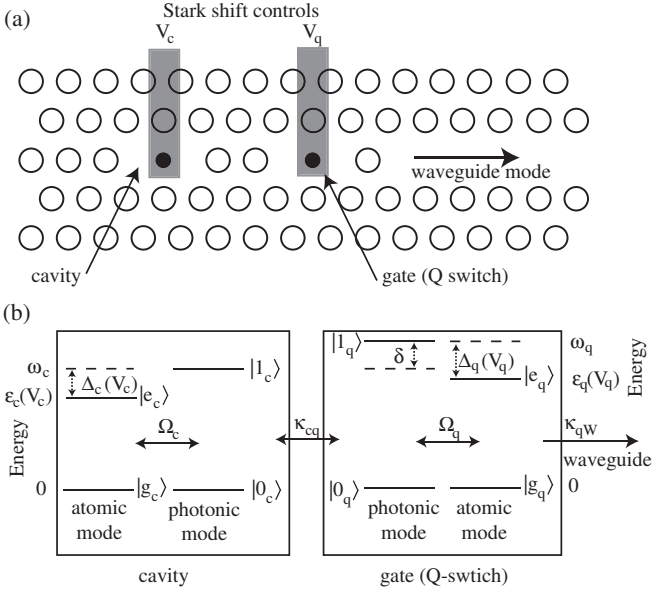}
\caption{\label{fig:CoupledPBGCavities} (a) Schematic of coupled
cavity system.  The open holes correspond to the PBG lattice, where
the missing holes constitute the cavities and waveguide.  The filled
circles are the atoms, with resonance controlled via top gates
(shaded rectangles).  The left defect is the `cavity', the right the
gate, and outcoupling is via the waveguide mode on the right. (b)
Diagram of states, energy levels, energy separations and couplings
in the bare basis.  Solid lines represent energy levels, dashed
lines are used to show the references used for energy separations
and arrows to depict coupling, reversible couple by double arrows,
irreversible coupling (outcoupling) via a single arrow. All
frequencies as shown in the diagram are positive.  After initial
fabrication, we only have control over the atomic transition
energies, $\varepsilon_c$ and $\varepsilon_q$.}
\end{figure}

The structure we consider is a coupled cavity arrangement, similar
in spirit to that studied by Waks and Vukovic \cite{bib:WaksOE2005},
where two defects in the PBG lattice were placed in close proximity
to form evanescently coupled cavities.  Our arrangement is shown
schematically in Fig~\ref{fig:CoupledPBGCavities} (a), where the
left-hand cavity is the storage cavity (or simply cavity), the right
hand cavity is the gate, which is in close proximity to a waveguide,
or other leaky, classical region. In this limit we can describe the
coupling between the distinct regions (cavity, gate and waveguide),
which is due to evanescent leakage of the electromagnetic modes, as
being equivalent to photon hopping between the regions
\cite{bib:PhotonHopping}. In addition to the previously considered
systems, however, we augment this arrangement by placing a single
two-state atomic system in the center of the cavity and gate
\cite{bib:BadolatoScience2005}, where the transition frequency of
the atom can be controlled by some external control field. An
example of a system that could realize such an architecture would be
a single crystal diamond with photonic crystal drilled using
focussed ion beam milling and liftoff \cite{bib:OliveroAdvMat2005}
where single ion implantation techniques \cite{bib:JamiesonAPL2005}
are used to locate individual nitrogen-vacancy centers in the centre
of the cavity, controlled via the linear Stark shift
\cite{bib:RedmanJOSAB1992}.  It is this control of the atomic
frequency that constitutes our sole \emph{dynamic} (i.e.
post-fabrication) control of the system parameters, and is
responsible for the Q-switching possibilities that we discuss in
this paper.

The method for Q switching this system can be understood easily, and
is a logical extension of previous work on cavity QED and photon
blockade \cite{bib:PhotonBlockade}. Firstly the cavity is arranged
so that one and only one photon is loaded into the cavity via some
external pump, and the gate is in its ground state. The cavity and
gate resonances are initially dissimilar, so that light from the
cavity cannot leak across to the gate.  Secondly the eigenmodes of
the gate are varied by changing the resonance frequency of the atom
in it, and when one of the gate modes is resonant with a mode of the
cavity, photon hopping occurs. The gate is a relatively bad cavity,
coupled to the output modes of a waveguide, and so photons leaking
into the gate are rapidly outcoupled to the waveguide.  As photon
hopping is the source of the cavity-gate coupling, it is clear that
optimal outcoupling results from balancing the competing needs of
large cavity-gate detuning, with photonic population of the resonant
mode of the gate at the outcoupling resonance.  These points will be
made more explicit by considering the model Hamiltonian.

The Hamiltonian for our system is written
\begin{eqnarray}
\mathcal{H} &=& \mathcal{H}_c + \mathcal{H}_{q} +
\mathcal{I}_c+ \mathcal{I}_{q} + \mathcal{P} , \label{eq:Ham} \\
\mathcal{H}_{\alpha}/\hbar &=& \varepsilon_{\alpha} (V_{\alpha})
\sigma_{\alpha}^-
\sigma_{\alpha}^+ + \omega_{\alpha} a_{\alpha}^{\dag} a_{\alpha}, \\
\mathcal{I}_{\alpha}/\hbar &=& \Omega_{\alpha}
\left(\sigma_{\alpha}^- a_{\alpha} + \sigma_{\alpha}^+
a_{\alpha}^{\dag}\right), \\
\mathcal{P}/\hbar &=& \kappa_{cq} \left(a_{q}^{\dag} a_c + a_{q}
a_c^{\dag} \right),
\end{eqnarray}
where $\mathcal{H}_{\alpha}$ and $\mathcal{I}_{\alpha}$ refer to the
bare and interacting parts of the Hamiltonians respectively, for
$\alpha = c,q$ for cavity or gate (Q-switch).
$\varepsilon_{\alpha}(V_{\alpha})$ is the transition frequency of
the atom in $\alpha$ which can be controlled by the Stark shifting
gates at some potential $V$, the exact functional dependance of the
Stark shift on gate potential is not important. $\omega_{\alpha}$ is
the resonance frequency of the photon in $\alpha$, and
$\Omega_{\alpha}$ is the atom-cavity coupling (one-photon Rabi
frequency) in $\alpha$. The $\sigma_{\alpha}$ are the usual Pauli
operators for the atoms in $\alpha$, and $a_{\alpha}$ is the usual
photon annihilation operator in $\alpha$. $\mathcal{P}$ describes
the photon hopping, with coupling $\kappa_{cq}$.  Coupling to the
external waveguide is described via a non-Hamiltonian term which
will be introduced in the density matrix formalism.  All these terms
are depicted schematically in Fig.~\ref{fig:CoupledPBGCavities} (b).

In general, the two cavity system with two atoms is a moderately
complicated problem to treat exactly, however by considering just
the one quantum manifold (i.e. where only one quantum of excitation
is in the system), and assuming that the detuning between the
cavities is large, i.e. $\omega_{q} - \omega_c = \delta \gg
\kappa_{cq},\Omega_{\alpha}$, we get significant insight.  In this
limit we can solve each cavity independently (i.e. ignoring
$\kappa_{cq}$ as our zeroth order approximation) to get the
approximate eigenstates, which are the well-known dressed states,
\begin{multline}
|\pm_c g_{q} 0_{q}\rangle = \frac{\left(-\frac{\Delta_c (V_c)}{2}
\pm \chi_c\right) |g_c 1_c\rangle +  \Omega_c |e_c 0_c
\rangle}{\sqrt{2\chi_c^2 \pm
\chi_c\Delta_c (V_c)}} |g_{q} 0_{q}\rangle, \\
|g_c 0_c \pm_{q}\rangle = \frac{\left(-\frac{\Delta_{q}(V_q)}{2} \pm
\chi_{q}\right) |g_{q} 1_{q}\rangle + \Omega_q |e_{q} 0_{q}\rangle
}{\sqrt{2\chi_{q}^2 \pm \chi_{q}\Delta_{q}(V_q)}} |g_c 0_c\rangle,
\end{multline}
where we have introduced $|g\rangle$ and $|e\rangle$ as the states
of the atoms, $\Delta_{\alpha}(V_{\alpha}) = \omega_{\alpha} -
\varepsilon_{\alpha} (V_{\alpha})$, the detuning, and $\chi_{\alpha}
= \sqrt{[\Delta_{\alpha}(V_{\alpha})/2]^2 + \Omega_{\alpha}^2}$, the
generalized Rabi frequency.  The associated eigenenergies are
\begin{eqnarray}
E_{|\pm_c,g_{q} 0_{q}\rangle} &=& \pm \chi_c -\Delta_c(V_c)/2 , \\
E_{|g_c,0_c \pm_{q}\rangle} &=& \delta \pm \chi_{q} -
\Delta_{q}(V_q)/2 .
\end{eqnarray}

By setting $\Delta_c = 0$ and $\Delta_{q} = -\delta < 0$, we can
calculate the approximate interaction strength (coupling matrix
element) of the gate induced resonance between the cavities, which
is (for example between $|+_c g_{q} 0_{q} \rangle$ and $|g_c 0_c
-_{q}\rangle$ at the gate defined resonance $\Delta_{q} = -\delta +
\Omega_c$)
\begin{eqnarray}
\mathcal{J} &=& \langle g_c 0_c -_{q}|\mathcal{P} |+_c g_{q}
0_{q}\rangle = \frac{1}{\sqrt{2}}\frac{\Omega_q}{\delta}\kappa_{cq}.
\label{eq:JRes}
\end{eqnarray}
The value of $\mathcal{J}$ sets the time-scale for the interaction
photon hopping.  In particular, if we wish to adiabatically transfer
the excitation from the cavity to the gate, the sweep rate of the
gate should be slow compared to $1/\mathcal{J}$.

\begin{figure}[tb!]
\includegraphics[width=0.8\columnwidth,clip]{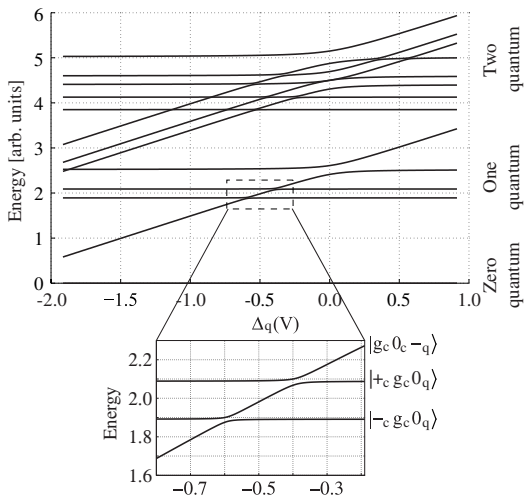}
\caption{\label{fig:CCEV} Eigenvalues for the two-cavity system as a
function of $\Delta_{q}$ for $\Omega_c = \Omega_q = \kappa_{c q} =
0.1$, $\omega_c = 2$, $\omega_{q} = 2.5$ and $\delta=0.5$. The
eigenspectrum naturally divides into the component manifolds, we are
most interested in the one quantum manifold.  The bottom trace shows
a closeup of the one quantum manifold, highlighting interactions
between the cavities, via the off-resonant dressed state.  The
resultant anti-crossings indicate coupling between the gate and
cavity, and hence where switching can occur.}
\end{figure}

To further explore the coupling between the cavities, we present in
Fig.~\ref{fig:CCEV} the eigenspectra determined by numerically
solving the Hamiltonian in Eq.~\ref{eq:Ham} without further
approximation as a function of $\Delta_{q}(V_q)$, for $\Omega_c =
\Omega_q = \kappa_{c q} = 0.1$, $\omega_c = 2$, $\omega_{q} = 2.5$
and $\delta = \omega_{q}-\omega_c = 0.5$.  These parameters were
chosen to simply highlight the important system features. The system
conveniently breaks into three manifolds, distinguished by the total
number of quanta, the zero quantum manifold is the lowest, along the
$\Delta_q(V_q)$ axis, then the one quantum and two quantum.  As we
are most interested in the resonances between the cavity and gate in
the one quantum manifold, we present a closeup of this in the inset
to Fig.~\ref{fig:CCEV}, where the anti-crossings indicating photon
hopping between the cavity and gate are clearly visible.  Note that
these parameters were merely chosen to demonstrate the relevant
processes, and all units are arbitrary.

\begin{figure}[tb]
\includegraphics[width=0.8\columnwidth,clip]{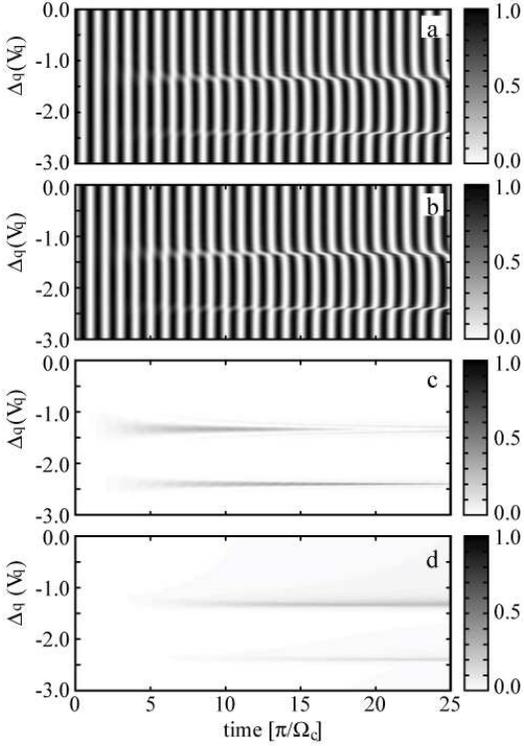}
\caption{\label{fig:TransEvol} Transient evolution of the
cavity-gate-waveguide system showing populations in the states (a)
$|g_c 1_c g_{q} 0_{q} 0_W\rangle$, (b) $|e_c 0_c g_{q} 0_{q}
0_W\rangle$, (c) $|g_c 0_c g_{q} 1_{q} 0_W\rangle$ and (d) $|g_c 0_c
g_{q} 0_{q} 1_W\rangle$ respectively (the population in $|g_c 0_c
e_{q} 0_{q} 0_W\rangle$ is never visible for these parameters), as a
function of time and $\Delta_{q}$ given initial state $|g_c 1_c
g_{q} 0_{q} 0_W\rangle$ for $\Omega_c = \Omega_{q} = 0.5$,
$\kappa_{cq} = \kappa_{q m} = 0.1$ and $\delta = 2$.  Note the
oscillation frequency in the range $-\delta - \Omega_c < \Delta_q <
-\delta + \Omega_c$ is found to be $\sqrt{\Omega_c^2 +
\mathcal{J}^2}$.  The maximum of the population transfer peaks in
the waveguide in (d) are $0.26$ at $\Delta_q \sim -\delta +
\Omega_c$, and $0.17$ at $\Delta_q \sim -\delta - \Omega_c$.}
\end{figure}

The previous analysis just treats coupling between the cavity and
gate, but to proceed further we need to include the coupling to the
waveguide mode ($W$).  This is best done by introducing an
irreversible loss term, analogous to spontaneous emission, which
models coupling into an extra waveguide mode.  We then solve the
density matrix equations of motion to examine the transient coupling
into the waveguide mode. Concretely we solve for the density matrix,
$\rho$, using:
\begin{multline}
\dot{\rho} = -\frac{i}{\hbar} [\mathcal{H},\rho] + \kappa_{q w}
\mathcal{L}[\rho,a_W^{\dag}a_{q}], \\
\mathcal{L}[\rho,a_W^{\dag}a_{q}] = a_W^{\dag}a_{q} \rho
a_q^{\dag}a_{W} - \frac{a_q^{\dag}a_{W} a_W^{\dag}a_{q} \rho + \rho
a_q^{\dag}a_{W} a_W^{\dag}a_{q}}{2}.
\end{multline}
$\kappa_{q W}$ is the gate-waveguide coupling. An example of the
evolution obtained is presented in Fig.~\ref{fig:TransEvol} which
shows the populations in the bare state basis, from top to bottom:
$|g_c 1_c g_{q} 0_{q} 0_W\rangle$, $|e_c 0_c g_{q} 0_{q}
0_W\rangle$, $|g_c 0_c e_{q} 0_{q} 0_W\rangle$, $|g_c 1_c g_{q}
0_{q} 0_W\rangle$ and $|g_c 0_c g_{q} 0_{q} 1_W\rangle$
respectively, as a function of time and $\Delta_{q}(V_q)$ given
initial state $|g_c 1_c g_{q} 0_{q} 0_W\rangle$. Clearly noticeable
are coherent oscillations corresponding to Rabi oscillations in the
cavity, and the gradual buildup of population in the waveguide mode.
The increase in the oscillation frequency in the region
$-\delta-\Omega_c < \Delta_q < -\delta+\Omega_c$ is a consequence of
an increased eigenvalue splitting, similar to (but more complicated
than) that seen in multiply coupled three state systems, see for
example Ref.~\onlinecite{bib:GreentreeAT}.  In the simpler,
doubly-driven three-state case, the oscillation frequency is given
by the sum of the squares of the Rabi frequencies of the driving
fields.  In this case, the result is similar, with the oscillation
frequency given by the sum of the squares of the interaction matrix
elements, i.e. $\sqrt{\Omega_c^2 + \mathcal{J}^2}$.

\begin{figure}[tb]
\includegraphics[width=0.8\columnwidth,clip]{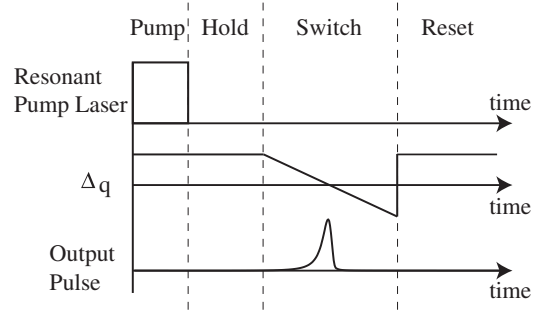}
\caption{\label{fig:SweepSteps} Steps required to adiabatically
generate Q-switched single photon pulses as a function of time.
Pump: first the cavity is pumped with a laser at frequency $\omega_c
+ \Omega_c$, and pulse area $\pi$, storing one quantum in the
cavity. Hold: the system is maintained for a length of time, set by
requirements of triggering the rest of the system and acceptable
loss probabilities.  Switch: $\Delta_q$ is adiabatically swept
through the resonance, and the single photon pulse outcoupled.
Reset: the system is returned to its initial state for repeated
operation.}
\end{figure}

Although illustrating much of the necessary physics, it is clear
that the results in Fig.~\ref{fig:TransEvol} do \emph{not}
illustrate an effective mechanism for single photon generation.  The
reason here is that the outcoupling probability can be no better
than $50\%$, and the Rabi oscillations render the system prone to
nonadiabatic errors. Also pertinent is that a complicated set of
interference fringes are observed which need to be understood for
transient analysis. Given such limitations, it is preferable to
initialize the system in the state $|+_c g_{q} 0_{q} 0_W\rangle$ and
follow an adiabatic transition along the anticrossing between $|+_c
g_{q} 0_{q} 0_W\rangle$ and $|g_c 0_c -_{q} 0_W\rangle$. By ensuring
that the gate and waveguide mode are strongly coupled,  i.e.
$\kappa_{qW}$ is large compared with the coupling matrix elements,
population in the state $|g_c 0_c -_{q} 0_W\rangle$ will be rapidly
transferred to $|g_c 0_c g_{q} 0_{q} 1_W\rangle$, and hence the
cavity-gate resonance will act as an effective Q-switch for the
cavity. Initialization of the system could be achieved by pumping
the cavity with light of frequency $\omega_c + \Omega_c$, which
would be resonant with the $|g_c 0_c\rangle - |+_c\rangle$
transition, but not resonant with transitions to the two quanta
manifold.  A schematic of the steps required to adiabatically
outcouple the single photon is shown in Fig.~\ref{fig:SweepSteps}

\begin{figure}[tb!]
\includegraphics[width=0.8\columnwidth,clip]{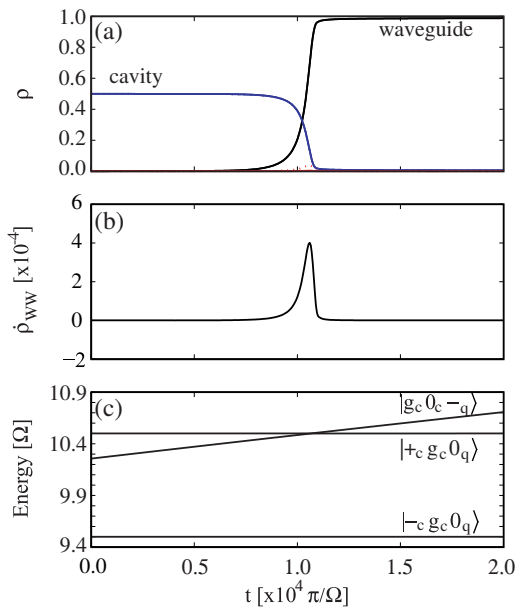}
\caption{\label{fig:AdiabaticPulse} (a) State population (expressed
in the bare basis) during the adiabatic sweep as a function of time
given initialization in the state $|+_c g_q 0_q\rangle$ at $t=0$.
Observe the smooth variation in population from cavity superposition
states to the waveguide mode with minor, transient occupation of the
gate. This is slightly asymmetric due to the asymmetric coupling to
and from the gate. (b) Time derivative of waveguide occupation,
$\dot{\rho}_{WW}$, proportional to the output intensity.  Again, the
pulse is asymmetric, although a tailored bias sweep could symmetrize
the pulse.  (c) Eigenvalues during the sweep, note that on this
scale, the avoided crossing between $|+_c g_q 0_q\rangle$ and $|g_c
0_c -_q\rangle$ is not resolved.  For this example, the parameters
chosen were $\Omega_c = \Omega_{q} = \Omega$, $\delta = 4\Omega$,
$\kappa_{cq} = 0.01 \Omega, \kappa_{q W} = 0.1 \Omega$, $-3.2 \Omega
\leq \Delta_{q} \leq -2.2 \Omega$ and the total sweep length was
$T_{\max} = 2 \times 10^4 \pi/\Omega$}
\end{figure}

The results of the adiabatic transfer from cavity to gate are shown
in Fig.~\ref{fig:AdiabaticPulse}, clearly showing both the
population in $|g_c 0_c g_{q} 0_{q} 1_W\rangle$, which we denote
$\rho_{WW}$, and the time derivative of this population,
$\dot{\rho}_{WW}$, which is proportional to the intensity of the
resulting photon pulse.  In this case we chose $\Omega_c =
\Omega_{q} = \Omega$, $\delta = 4\Omega$, $\kappa_{cq} = 0.01
\Omega, \kappa_{q W} = 0.1 \Omega$, $-3.2 \Omega \leq \Delta_{q}
\leq -2.2 \Omega$ and the length of the sweep was $T_{\max} = 2
\times 10^4 \pi/\Omega$. Note that because of the difference between
$\kappa_{c q}$ and $\kappa_{q W}$ the resultant single photon pulse
is not a Gaussian.  To retrieve a Gaussian pulse, one could either
choose a system with equal photon hopping matrix elements or a more
complicated gate sweep.  Note that the integral of the derivative is
unity, as required for a pulse of one photon.

When considering the operating parameters of the Q switch, it is
also necessary to determine the quiescent fidelity,  i.e. the photon
leakage from cavity to the Q-switching gate when the switch is not
activated. For simplicity, if we assume $\Delta_c = \Delta_{q} = 0$
and $\delta \gg \kappa_{c q}, \Omega_c, \Omega_{q}$, then the
population in state $|+_c g_q 0_q\rangle$, $\rho_{+_c}(t)$ at time
$t$, given initialization in state $|+_c g_q 0_q\rangle$ at $t=0$,
is
\begin{eqnarray}
\rho_{+_c} (t) = \exp\left(-\frac{\kappa^2_{cq}}{2 \delta^2}
\kappa_{qW} t\right)
\end{eqnarray}
where $\kappa_{cq}^2/\delta^2$ is the standard, steady state,
off-resonant population leaking from the cavity to gate, which is
then outcoupled at a rate $\kappa_{qW}$.  Under the conditions used
to generate Fig.~5, this equates to a population of $\rho_{+_c}=
0.98$ at $t=2\times 10^4 \pi/\Omega$, or alternatively, at worst a
$2\%$ probability of the photon outcoupling from the cavity.

Finally we comment on the practicality of realizing our scheme in a
realistic structure, and for our purposes we assume a PB cavity
structure fabricated in diamond containing a single NV$^-$ centre at
the maximum of the cavity mode.  The wavelength of the zero-phonon
line resonance of an NV$^-$ centre is $\lambda = 638 \times 10^{-9}
\mathrm{m}$, with frequency $\omega = 2.95 \times 10^{15}
\mathrm{Hz}$, and assuming that each cavity has volume $V =
\lambda^3 = (638\times 10^{-9})^3 \mathrm{m}^3$, then the atom
cavity coupling will be $\Omega = \mu \sqrt{\omega/(2 \hbar
\epsilon_0 V)} \sim 10^{10} \mathrm{Hz}$ (given the electric dipole
moment of the NV$^-$ centre of $\mu \sim 10^{-29} \mathrm{C
m}^{-1}$).  For this degree of coupling, the tuning range of the
centres should be many $\Omega$. The tuning range reported in Ref.
\onlinecite{bib:RedmanJOSAB1992} is $\sim 10^{12} \mathrm{Hz}$,
which does not constitute an upper limit on the Stark tuning.
Therefore the atomic tuning criterion should be easy to satisfy, and
we presume $\delta = 10^{12}\mathrm{Hz}$.

If we assume that the cavities are in the good cavity limit, and
that the cavity Q is dominated by photon loss due to the photon
hopping between cavities and the waveguide, then the cavity Q must
be fairly large to ensure minimal population leakage when we are not
at the switching point. The figure of merit here is that the ratio
$\kappa_{cq}^2/\delta^2$ should be small.  If we aim for a residual
population of $10^{-4}$, then $\kappa_{cq}/\delta \leq 10^{-2}$,
i.e. $\kappa_{cq} = 10^{10}\mathrm{Hz}$, and $\kappa_{qW} = 10
\kappa_{cq} = 10^{11}\mathrm{Hz}$. If we assume that the cavity Q is
dominated by the photon hopping terms, then we have (for the cavity)
\begin{eqnarray}
Q_c = \frac{\omega}{\kappa_{cq}} \sim 10^5
\end{eqnarray}
and the Q of the gate will be $10^4$.  Although technically
demanding, we note that Q factors $\sim 10^7$ have been shown to be
possible in silicon photonic bandgap cavities on silica
\cite{bib:SongNM2005}. Furthermore, although we have studied our
device in this demanding regime to clarify the effects, proof of
principle experiments will be possible with significantly lower Q
values by relaxing the requirements for adiabatic transfer and
increasing the detuning of the cavities.

With these parameters, the pulse obtained by adiabatically switching
the gate will be outcoupled in a time commensurate with
$1/\mathcal{J} = 10^{-9} \mathrm{s}$. Without switching, the
expected population in the waveguide mode over this timescale would
be $0.01$.

The full set of required parameters are summarized in
Table~\ref{tab:NVParms}.

\begin{table}[tb!]
\begin{tabular}{|l|c|}
\hline \hline Parameter & Value \\
\hline \hline
Wavelength & 638 $\mathrm{nm}$ \\
\hline Transition frequency & $2.95 \times 10^{15} \mathrm{Hz}$ \\
\hline $\Omega_c = \Omega_q$ & $10^{10} \mathrm{Hz}$ \\
\hline $Q_c$ & $10^5$ \\
\hline $Q_q$ & $10^4$ \\
\hline $\kappa_{cq}$ & $10^{10}\mathrm{Hz}$ \\
\hline $\kappa_{qw}$ & $10^{11}\mathrm{Hz}$ \\
\hline $\delta$ & $ \leq 10^{12}\mathrm{Hz}$ \\
\hline \hline
\end{tabular}
\caption{\label{tab:NVParms} Nominal parameters for efficient
Q-switching for NV$^-$ centres embedded in an all-diamond photonic
crystal.}
\end{table}

In conclusion, we have presented a scheme for Q switching a photonic
bandgap cavity by controlling the resonance condition of an adjacent
cavity. Each cavity contains a single two-level atom, and the
transition frequency of the atom can be controlled via a Stark
shifting electrode.  We refer to the right-hand cavity as the gate
which Q-switches the cavity. The resonance frequencies of the two
cavities are initially dissimilar, but by tuning the atomic
transition in the gate, a resonance condition between the cavity and
gate is obtained, resulting in photon hopping between the cavities.
By introducing a waveguide mode adjacent to the gate, photons leak
rapidly out of the gate. Such a device constitutes a solid-state
source of transform limited single photons on demand. An ideal
system to test such concepts would be in micromachined diamond
containing the nitrogen-vacancy color centre, although our ideas can
be applied to any photonic bandgap cavity containing a two-level
atom in the maximum of the cavity mode.

ADG would like to acknowledge useful discussions with T. Ralph, and
J. Cole. JS acknowledges support from the Fund for Promotion of
Research at the Technion.  This work was supported by the Australian
Research Council, the Australian government and by the US National
Security Agency (NSA), Advanced Research and Development Activity
(ARDA) and the Army Research Office (ARO) under contracts
W911NF-04-1-0290 and W911NF-05-1-0284.

%%%%%%%%%%%%%%%%%%%%%%%%%%%%%%%%%%%%%%%%%%%%%%%%%%%%%%%%%%%%%%%%

\end{document}